**Self-assembly and electron-beam-induced direct etching of suspended graphene nanostructures**

Sarah Goler[1,2], Vincenzo Piazza[2], Stefano Roddaro[1], Vittorio Pellegrini[1], Fabio Beltram[1,2], and Pasqualantonio Pingue[1,*]

*1) Laboratorio NEST - Scuola Normale Superiore, and Istituto Nanoscienze - CNR, Piazza San Silvestro 12, I-56127 Pisa, Italy*

*2) Center for Nanotechnology Innovation @ NEST, Istituto Italiano di Tecnologia, Piazza San Silvestro 12, 56127 Pisa*

May 2011

## ABSTRACT

We report on suspended single-layer graphene deposition by a transfer-printing approach based on polydimethylsiloxane stamps. The transfer printing method allows the exfoliation of graphite flakes from a bulk graphite sample and their residue-free deposition on a silicon dioxide substrate. This deposition system creates a blistered graphene surface due to strain induced by the transfer process itself. Single-layer-graphene deposition and its "blistering" on the substrate are demonstrated by a combination of Raman spectroscopy, scanning electron microscopy and atomic-force microscopy measurements. Finally, we demonstrate that blister-like suspended graphene are self-supporting single-layer structures and can be flattened by employing a spatially-resolved direct-lithography technique based on electron-beam induced etching.

**I. INTRODUCTION**

Graphene is a single atomic layer of carbon atoms arranged in a honeycomb lattice. This material is at the moment of much interest both for fundamental studies and in light of electronic and photonic applications since it hosts a high-mobility two-dimensional (2D) electron/hole gas displaying rather peculiar properties [1,2,3]. Many experimental studies were made possible by a simple micromechanical cleavage method employing scotch tape [1,4] thanks also to rather straightforward methods allowing the identification of actual graphene (i.e. monoatomic) samples, e.g. by optical microscopy on a silicon oxide ($SiO_2$/Si) substrate [5]. The 2D nature of graphene is clearly manifested in Raman studies that exhibit significant differences with respect to graphite or multi-layer graphene flakes [6]. Transport and optical properties of graphene are rsther unique as a result of its peculiar band structure with zero gap and linear dispersion of conduction and valence bands at the corners of the Brillouin zone (Dirac points) [1,2,3]. Further control on these properties can be achieved by tuning sample strain [7], allowing for example the creation of a band-gap due to sublattice-symmetry breaking [7,8,9,10]. In particular, the presence of nanometer-scale corrugations was predicted to be of interest for hydrogen-storage applications [11].

This remarks suggest that fabrication of graphene on large areas with the controlled definition of local strain patterns us a challenge that may be of much importance for the development of graphene-based nano-electronics or nano-photonics. To this end there has been significant progress in the fabrication of atomic layers of graphene epitaxially grown on silicon- and carbon-terminated faces of silicon carbide [12,13,14], where strain can result from the interaction with the substrate, and on metallic surfaces as Ni and Cu [15], where the presence of a rippled structure was demonstrated by scanning tunneling microscopy (STM) [16]. Yet much remains to be done to achieve graphene layers of quality comparable to that made available by simple micromechanical cleavage.

Recent studies showed that transfer-printing approaches are a valid technique for depositing high-quality graphene on various substrates and may represent a useful new route towards the development of graphene-based integrated circuits. These studies used resin [17], poly(methyl methacrylate) (PMMA) [18] or polydimethylsiloxane (PDMS) [19, 20]-based transfer protocols to fabricate high-quality thin-graphite or actual graphene samples.

In this letter, we employ a transfer-printing approach based on PDMS stamps, and show that it allows repeated depositions of single-layer graphene flakes on $SiO_2$/Si substrates with strain-induced blistering of their surface. Raman scattering, scanning-electron microscopy (SEM), atomic-force microscopy (AFM), and optical microscopy demonstrated both the achievement of the actual single-layer limit and the high structural quality of the graphene flakes as well as the presence of blisters on the graphene surface. Importantly PDMS is optically transparent and leaves no residues on the substrate contrary to resins or scotch tape. Moreover, it can also be used to stamp metallic pads for electrical contacts even with submicron resolution. For these reasons this method holds promises for the fabrication of complex circuits on graphene with the entire lithographic process being executed using PDMS stamps [21]. In this paper we show that due to the elastomeric characteristics of the PDMS rubber strain can be induced in the deposited graphene layers and that this strain relaxes locally leading to the formation of blisters on the graphene surface. Spatially resolved lithographic technique exploiting electron-beam-induced etching on graphene allow us to demonstrate that these blisters are suspended graphene membranes and that they relax their strain when pricked. Strain can also be present during the deposition process obtained by the standard scotch-tape method or in suspended-graphene layers [22] and it does have an impact on the resulting electrical and optical properties.

PDMS stamps were fabricated by a 10:1 prepolymer/initiator mixture. The viscous solution was mixed vigorously and placed in a vacuum desiccator for approximately 45 minutes to remove trapped-air bubbles. The PDMS was then poured onto a three-inch-diameter optically-polished silicon wafer, placed in a Petri dish and again in the vacuum desiccator to remove any remaining bubbles. This procedure yields stamps with a very flat surface. Final curing was then performed at $70^oC$ for 2 hours.

Oxygen-plasma treatment of the PDMS surface favors the graphite adhesion process described above. In order to produce plasma-treated surfaces, the silicone elastomer was placed in oxygen plasma for 20 seconds at 100 W and 80 mTorr. This process modifies the surface structure of the polymer. The details of the alterations of the surface are not fully understood, but it was proposed that negatively-polarized molecular groups are created that can strongly interact with polar molecules such as water [23]. Oxygen-treated PDMS stamps can also be used to remove part of the remaining small particles littering the surface of the $SiO_2$ substrate or to exfoliate previously-deposited multilayer graphene films directly off the substrate.

Our deposition process starts by placing and lightly pressing for a few minutes a square PDMS stamp (area 1 $cm^2$) on top of bulk highly-ordered pyrolytic graphite (HOPG) or natural graphite (pre-cleaned by cleaving the exposed surface with scotch tape). The stamp is then removed off the bulk graphite and a second stamp of PDMS is repeatedly put in contact with its graphite-/graphene-covered surface. This step was seen to increase the probability of actual single-layer graphene transfer when the PDMS stamp was placed in contact with a $SiO_2$/Si wafer for a few minutes, as usually done by employing the standard scotch-tape methodology.

As a result of this procedure graphene and multilayer graphite flakes are transferred to the $SiO_2$/Si substrate. Schematic drawings of the relevant steps of the present PDMS transfer-printing procedure are shown in Figs. 1(a)-(c). Figures 1(d) and (e) show that optical-microscope imaging can be employed to monitor in real time the transfer-printing procedure from the PDMS stamp to the $SiO_2$ surface thanks to the optical transparency of PDMS. In particular, Fig. 1(f) demonstrates a typical transfer-printing process on $SiO_2$ by optical microscopy imaging.

Initial classification of the number of graphene layers was carried out using an optical microscope under green-light illumination: this frequency range yields the best contrast for thin graphite on our $SiO_2$/Si substrates [5]. Raman spectroscopy and AFM imaging were then carried out to confirm the actual deposition of monolayer graphene as discussed below. The typical area of the graphene flakes obtained by the present PDMS transfer-printing method is about 10 $\mu m^2$, therefore large enough to allow further processing and the realization of electronic devices.

Figure 2(a) AFM (left) and scanning electron microscopy (SEM) (left) are images of a typical multi-layer graphene flake produced by this transfer method. In both images blistering is present on top of any small particle lying on the $SiO_2$ substrate. The height of these suspended membranes in this case is on the order of a few nanometers. Figure 2(b) shows another kind of blistering as visualized by AFM. In this case the rippling of graphene is much more pronounced and does not appear to be related to the presence of underlying nanoparticles. The typical height of the suspended membranes in this case was found of the order of tens of nanometers. Heating the sample at a temperature of 150 ºC for 20 minutes yielded a decrease of blister heights of less than 20% on average. In the case graphene-flake thickness greater than 10 nm, however, both density and height of blistering reduced drastically. The membranes were very robust and we were not able to rupture or prick blisters using the AFM tip in contact mode even employing very high forces (tens of μN range) [24, 25, 26].

The observed film structure can be linked to the pressure applied onto the PDMS stamp during its contact with the graphite in the first step of the transfer process. This pressure can cause the stretching of the PDMS surface in contact with the graphite surface. When the load is removed, the stamp relaxes to equilibrium causing the rippling of the attached graphite and graphene layers, as already observed for the case of thin metal layers on top of PDMS surface [27,28]. When they are transferred onto the substrate, the strained graphene layers relax in correspondence of particles present on the substrate surface or near depressions in the $SiO_2$ surface, where the van der Waals interactions are weaker [29].

In order to evaluate the structure and electronic properties of the transfer-printed graphene layers, spatially resolved Raman spectroscopy was performed at room temperature with the 488 nm line of an Argon laser. A 100x objective lens was used to focus the laser to a spot with a diameter of less than 500 nm and the resulting scattered light was dispersed by a single-grating spectrometer onto a Peltier-cooled CCD. Scans of the sample surface revealed the characteristic spectra of the $E_{2g}$-symmetry G band at around 1580 $cm^{-1}$ and the 2D band with a Raman shift corresponding to approximately 2700 $cm^{-1}$ for actual single-layer graphene (SLG). Figure 3(a) shows a CCD image of a sample containing single and multilayer graphene deposited by PDMS transfer-printing, while Fig. 3(b) reports the corresponding AFM image. The narrow, symmetric 2D peak at 2690 $cm^{-1}$, a characteristic feature of SLG, can be seen in Fig. 3(c), blue line [4,30,31]. The corresponding G peak displays an intensity of approximately one fourth that of the 2D peak, as expected [4]. We performed Raman analysis both on and off a blister (see the inset of Fig. 3(c)), obtaining the spectra reported in Fig. 3(c). Data show a red shift of the G mode of $\approx 1$ cm$^{-1}$ and of the 2D modes by as much as $\approx 2.5$ cm$^{-1}$. We recall that low-energy shifts of Raman lines were also observed in free-standing graphene [32] and were associated to a variation of electron (or hole) doping [28,33]. Following the previous studies on uniaxial strain induced in graphene layers deposited on poly(ethylene terephthalate) (PET) [34] or on PDMS [35] we could estimate that in the present case the induced strain in the blistered regions is the order of 0.1 %, at least with respect to the adjacent flat regions. We should like to stress that no qualitative differences emerge from the comparison of the Raman spectra of graphene obtained by PDMS with those produced by the standard micromechanical exfoliation method. The high quality of the single layer graphene produced by PDMS is supported by the absence of the disorder-induced D peak at ~1350 cm$^{-1}$.

In order to evaluate the impact of strain on the topography, high resolution AFM measurements in intermittent-contact mode were performed on the same SLG sample of Fig. 3. These data confirmed the existence of true single-layer graphene and also the blistering of the graphene thin layers, barely visible as a dark contrast in the optical microscopy images (Fig. 3(a)). In this sample, while some of the blisters are due to residual particles trapped beneath the membrane, in most cases these blisters seem to have a different origin (see AFM topography in Fig. 4 (a)). Phase imaging produces an increased contrast due to the different elastic behavior of blistered regions with respect to the flat areas (see Fig. 4 (b)). The flat regions are rippled (partially following the roughness of the substrate) and the elastic response of the blistered regions is lower (bright color) than that of the "flat" graphene and of the SiO$_2$ substrate (dark color scale). The largest blisters are located where small rigid debris are present (dark spot in phase image and bright one in topography). Smaller blistered structures can be observed even without a central debris suggesting that it may be the result of strain with no need for solid contaminants trapped underneath the membrane. To this end a high-resolution lithographic technique was carried out, based on electron-beam-induced etching (EBIE) of the graphene layer. The layer was imaged by a SEM with the capability to inject gases in the vacuum chamber itself (Merlin system by ZEISS). The same region of Fig. 4 was therefore employed to demonstrate that it is possible to perform a direct etching of graphene by injecting oxygen gas in the SEM chamber during imaging at low voltage (5 kV in this case) and exploiting the electron beam in order to locally create ozone and oxygen radicals [36]. We verified that the EBIE is particularly effective on SLG and we were able to monitor the etching process in real time by observing how the contrast changed during a line scan on the graphene flake. Figure 5(a) shows the SEM image of the graphene flakes together with the vertical scan line (dashed red line) where the single-line lithographic process was applied. Figure 5(b) reports the SEM image of the SGL after the spatially-resolved oxygen-radical etching. Two of the three larger blisters in Fig. 4(a) were strongly modified by the etching and the graphene membrane relaxed onto the surface. One of the blisters was pierced by direct EBIE (Fig. 5(c), red arrow). In this case the result was an almost complete flattening of the suspended SLG membrane (note the change in the dark image contrast between Fig. 5(a), (b) and Fig. 5 (c)). Strain relaxation probably originates from damages to the crystalline structure of graphene induced by oxygen. This interpretation was strengthened by the emergence of a strong D peak in the Raman spectra (data not shown) performed after the spatially-resolved oxygen radicals etching, in the vicinity of the exposed areas.

Blisters in graphene produced by standard micromechanical exfoliation (at much lower density compared to the present case) were observed and "bubbles" were also induced in exfoliated graphene by chemical methods and proton irradiation, as described in Ref. 26. Similar results on graphene or other single-layer materials obtained by PDMS-based transfer printing were reported: in those cases the presence of blistering was not highlighted by the authors but is evident from the published AFM images [37]. Recently, another group demonstrated that gold and silver nanoparticles deposited on top of $SiO_2$ substrates can create suspended graphene membranes [38], as we observed in correspondence of the nanoscale debris present on our substrates.

We believe that a more precise control of the pressure applied in the PDMS transfer-printing technique will favor the production of graphene layers having a surface topography ranging continuously from "flat" to "blistered" shapes.

## IV. CONCLUSIONS

In conclusion, PDMS transfer-printing can represents a very useful approach for the production of graphene from bulk graphite with much reduced contamination with respect to other transfer-printing methods that rely on scotch tape, resin or PMMA-based materials. We believe that it constitutes a straightforward methodology to investigate strain-induced effects in single-layer graphene. Moreover further improvements and optimization of the PDMS transfer-printing method reported here may allow the controlled and ordered self-assembly of blistering and rippling on single graphene layers on various substrates (e.g. as recently studied in suspended graphene membranes with thermally-generated strain [39, 40]), allowing a systematic investigation of the effect of blistering on the electronic and optical properties of graphene and their links and interactions with different substrate species. Finally, in order to prove the nature of the blistered nanostructures, a spatially resolved direct etching of graphene was successfully demonstrated employing e-beam induced ionization of $O_2$ in a SEM vacuum chamber. We think that this technique could become a valuable tool for graphene lithography once its impact on electronic and optical properties of the resulting patterned nanostructure is properly assessed [41].

**AKNOWLEDGEMENTS**

We would thank dr. Marco Cecchini for technical assistance during the PDMS sample preparation and functionalization.

**CAPTIONS**

**Figure 1.** Schematic drawings of the main steps for PDMS transfer-printing of graphene. (a) Lightly press the PDMS stamp on to the newly cleaved bulk graphite. Remove the stamp after a few minutes. (b) Press a clean PDMS stamp to the graphite- and graphene-covered stamp repeatedly. This process cleaves the graphite on the stamp, increasing the possibility of transferring a single layer. (c) Place the PDMS stamp on the $SiO_2$/Si wafer ensuring that the surface of the stamp is in contact with the wafer and leave in contact for a few minutes. (d) An image of graphite on PDMS as seen through an optical microscope. (e) Graphite on the $SiO_2$/Si substrate seen through the PDMS by an optical microscope. (f) Optical imaging of the transfer process from PDMS to the $SiO_2$ surface.

**Figure 2.** (a) AFM (left) and SEM (right) images of the same multi layer graphene flake on top of SiO2 substrate. Notice the blisters and the corresponding particles underneath. (b) AFM image of a rippled graphene structure before (left) and after (right) thermal annealing treatment.

**Figure 3.** (a) Optical image taken by 100x objective lens of single and multilayer graphene. The laser spot is situated on top of the monolayer of graphene. (b) AFM image of the same region. (c) Spatially-resolved Raman measurements performed in the area enclosed by the red box ("flat graphene") and in the blue one ("blistered graphene") as shown in the inset.

**Figure 4.** (a) Atomic-force microscopy topographic image of graphene. The bright spots reveal the blistering of the graphene layer. (b) Phase imaging of the same region, showing the high contrast between the blistered regions and the flat one due to a different elastic response. Notice

in some cases the presence of rigid structure (debris) in correspondence of the central part of the suspended membranes.

**Figure 5.** SEM images of the same SLG region. (a) Large vie of the sample: scale bar represents 2 μm, while the dashed red line is the one where EBIE was performed. (b) SEM picture of the SLG after the spatially resolved EBIE: the obtained cut has a width of 37 nm and some blisters disappeared, relaxing on the substrate. (c) The same region after EBIE pricking of the central blister (evidenced by the red arrow).

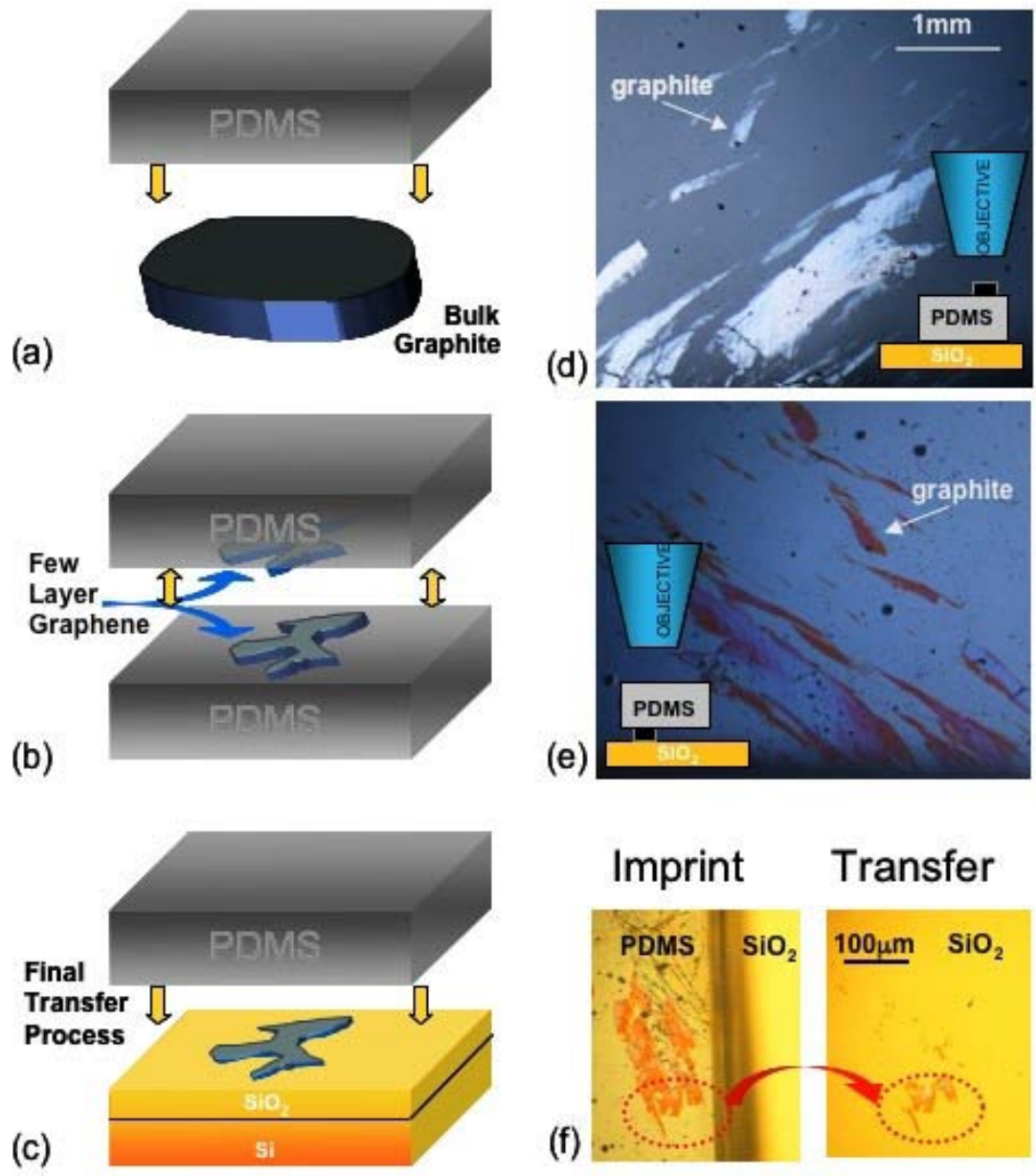

**FIGURE 1**

**(a)**        AFM                                                        SEM

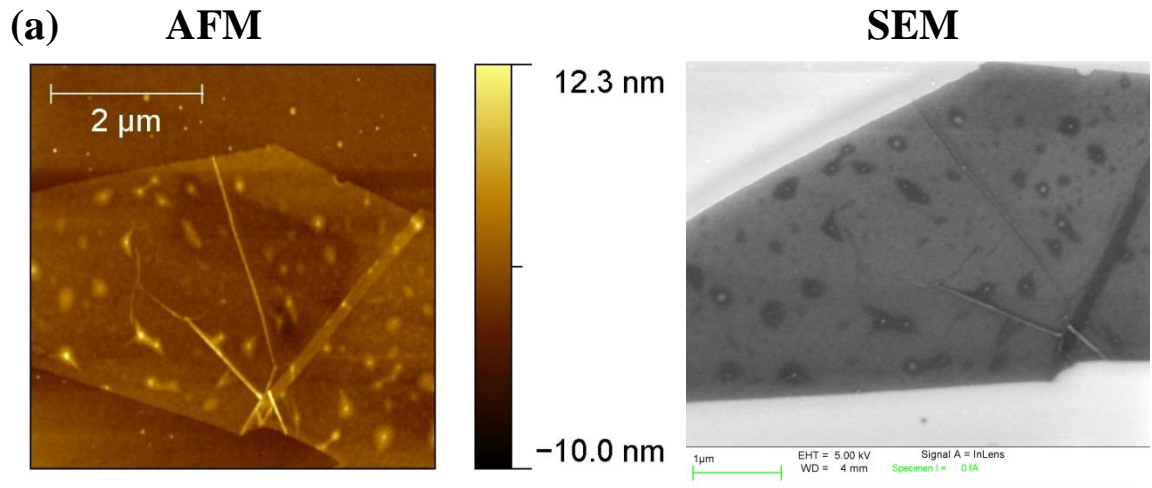

**(b)**   **BEFORE ANNEALING**                       **AFTER ANNEALING**

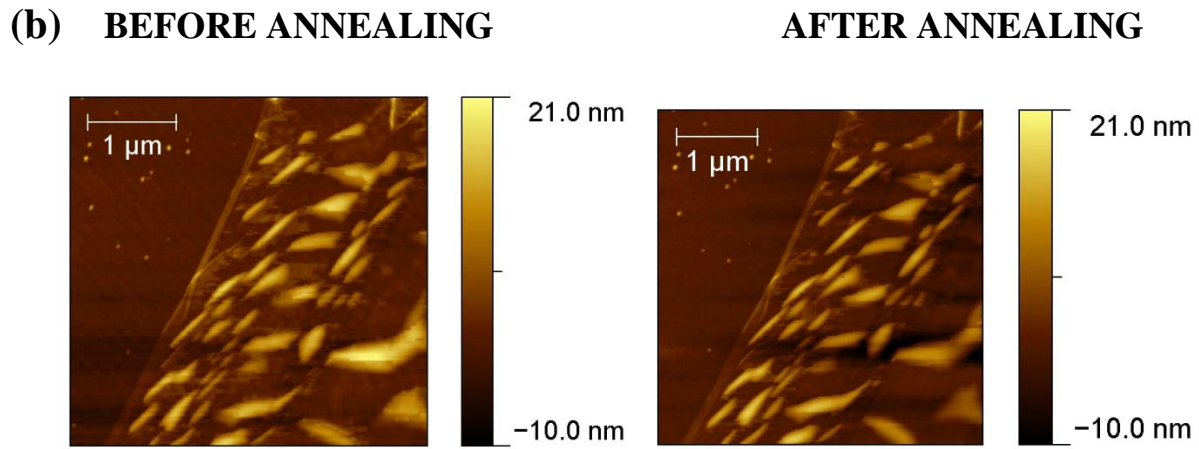

**FIGURE 2**

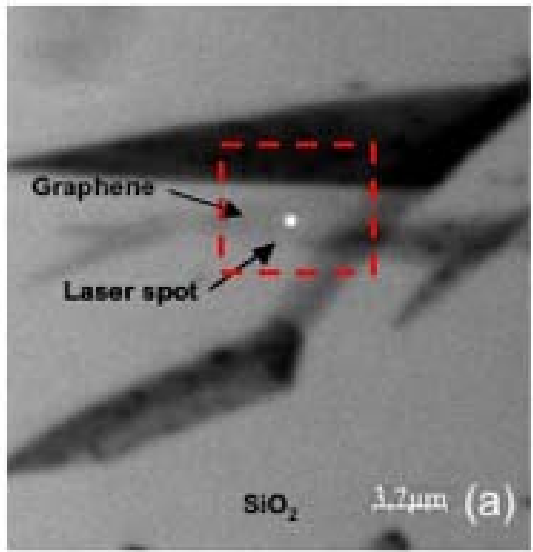
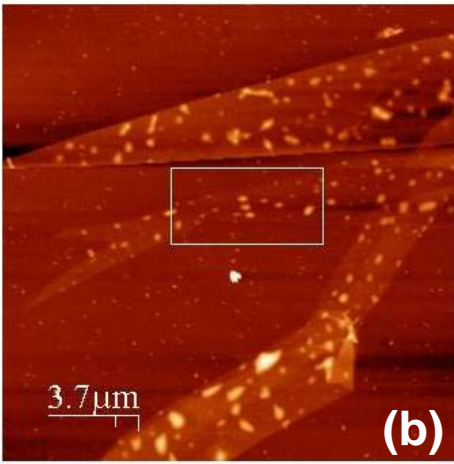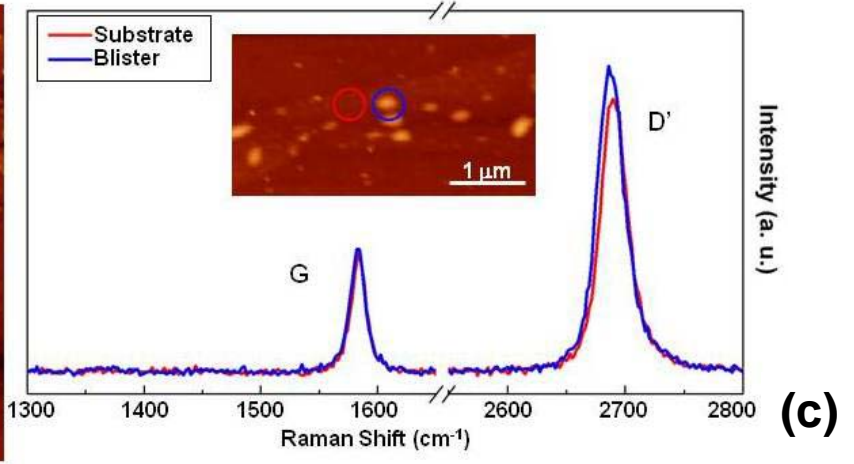

**FIGURE 3**

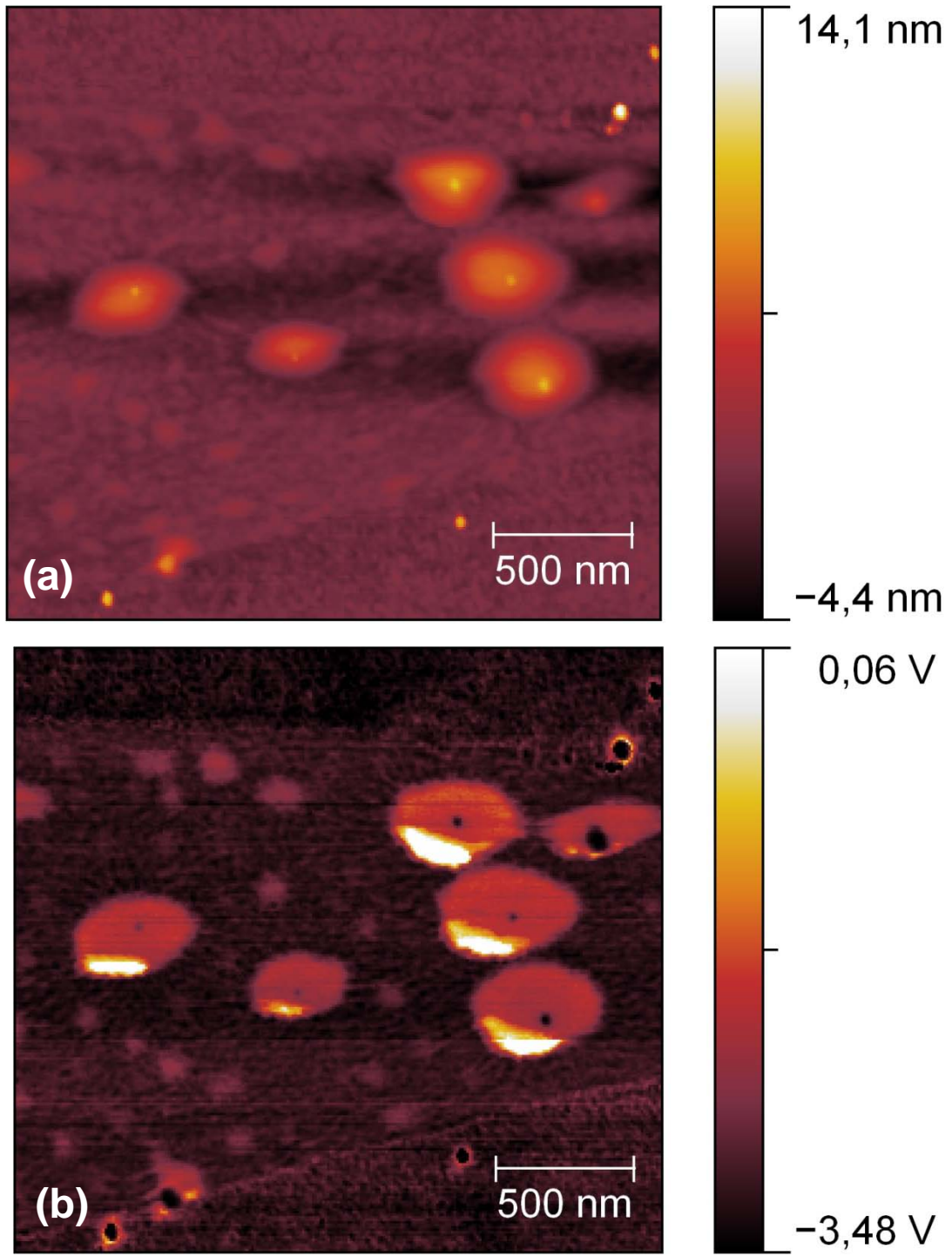

**FIGURE 4**

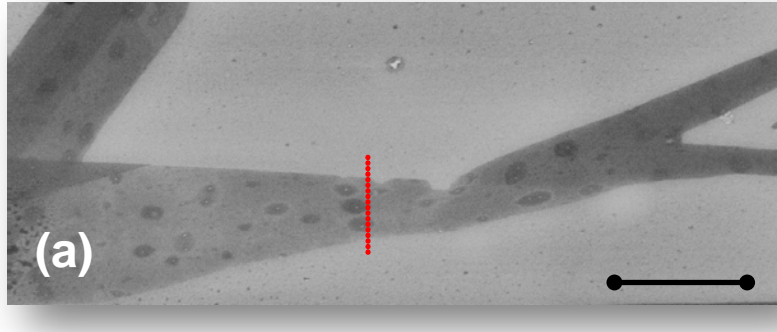
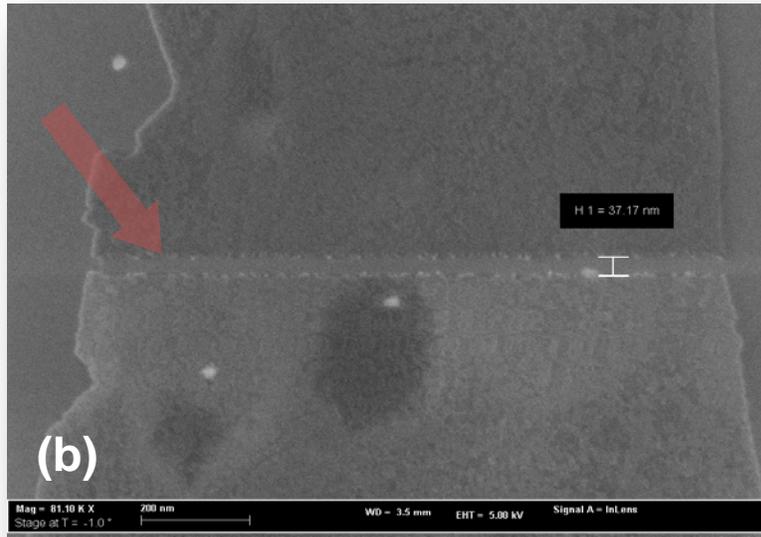
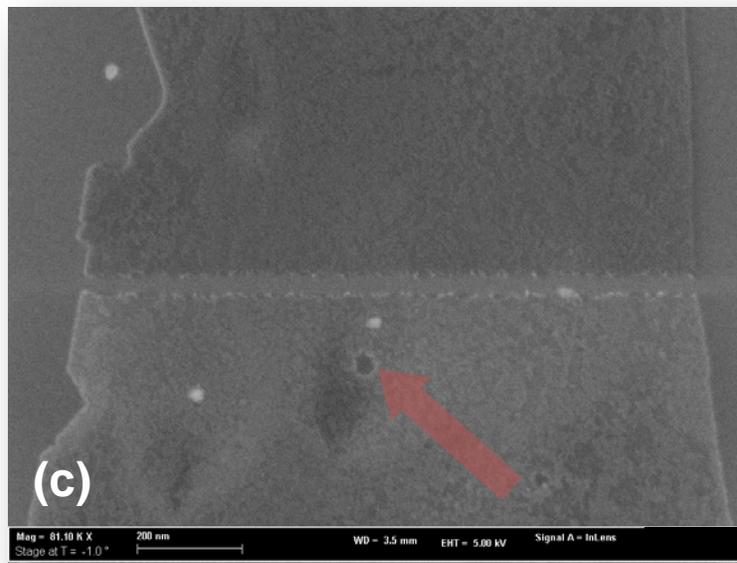

**FIGURE 5**